# Short- and medium-range orders in Al$_{90}$Tb$_{10}$ glass and their relation to the structures of competing crystalline phases

L. Tang[1], Z. J. Yang[1], T. Q. Wen[2], K. M. Ho[2,3], M. J. Kramer[2], and C. Z. Wang[2,3,*]

[1]*Department of Applied Physics, College of Science, Zhejiang University of Technology, Hangzhou, 310023, China*
[2]*Ames Laboratory-USDOE, Iowa State University, Ames, Iowa 50011, USA*
[3]*Department of Physics and Astronomy, Iowa State University, Ames, Iowa 50011, USA*

Corresponding author: * wangcz@ameslab.gov

**Abstract**

Molecular dynamics simulations using an interatomic potential developed by artificial neural network deep machine learning are performed to study the local structural order in Al$_{90}$Tb$_{10}$ metallic glass. We show that more than 80% of the Tb-centered clusters in Al$_{90}$Tb$_{10}$ glass have short-range order (SRO) with their 17 first coordination shell atoms stacked in a '3661' or '15551' sequence. Medium-range order (MRO) in Bergman-type packing extended out to the second and third coordination shells is also clearly observed. Analysis of the network formed by the '3661' and '15551' clusters show that ~82% of such SRO units share their faces or vertexes, while only ~6% of neighboring SRO pairs are interpenetrating. Such a network topology is consistent with the Bergman-type MRO around the Tb-centers. Moreover, crystal structure searches using genetic algorithm and the neural network interatomic potential reveal several low-energy metastable crystalline structures in the composition range close to Al$_{90}$Tb$_{10}$. Some of these crystalline structures have the '3661' SRO while others have the '15551' SRO. While the crystalline structures with the '3661' SRO also exhibit the MRO very similar to that observed in the glass, the ones with the '15551' SRO have very different atomic packing in the second and third shells around the Tb centers from that of the Bergman-type MRO observed in the glassy phase.

I.   **Introduction**

Rare-earth (RE) elements play an important role in modern materials sciences. While many RE containing compounds exist in oxide forms, there have been considerable interests in developing and understanding RE intermetallic compounds. Especially, aluminum-RE (Al-RE) alloys with high concentration of Al have shown to exhibit large strength-to-weight ratio and are promising for applications in many industrial areas [1-7]. However, the phase selection and competition in Al-rich Al-RE alloys are very complex. With more than 85 at. % Al and under rapid cooling, Al-RE alloys can form either metallic glass or metastable crystalline compounds, depending on the details of cooling procedures [8] as well as the introduction of other transition metals which would improve the glass formability [9-16]. While the glass formability range varies considerably with the RE elements, the best binary glass formers are Al-Sm and Al-Tb. This is quite interesting since the known stable intermetallic phase varies from $Al_{11}RE_3$ for the larger La-Sm to the $Al_3RE$ for smaller Y-Gd RE [8]. An overriding question regarding to this interesting phenomenon is how the structure of the amorphous or liquid state in these Al-RE systems affects the phase selection. Therefore, information about the short- to medium-range structural orders in these Al-RE alloys upon rapid cooling and their relationship to the local structure motifs in crystalline phases are vital to understanding the competition and phase selection between the glass formation and the crystallization. Recently, we have shown that $Al_{90}Sm_{10}$ undercooled liquid and glass exhibit strong SRO which are strongly correlated with the building blocks of several novel low-energy metastable crystalline structures in the vicinity of the $Al_{90}Sm_{10}$ composition [20-24]. However, none of these SRO motifs match well with the structures of the stable $Al_{11}RE_3$ or $Al_3RE$ phases. While many experimental efforts have also been devoted to investigating the structure of $Al_{90}Tb_{10}$ alloy [8, 25, 26], elucidating the detailed atomic structure in terms of SRO and MRO is still very challenging by experimental studies.

In this paper, using an accurate and reliable interatomic potential developed by an artificial neural network (ANN) deep learning method, we perform molecular dynamics (MD) simulations to study the structural order in $Al_{90}Tb_{10}$ metallic glass at the atomic level. Our MD simulations generated a $Al_{90}Tb_{10}$ glass sample which yields structure factor S(q) in very good agreement with the data from experimental X-ray diffraction (XRD) measurement. In order to investigate the structural order in the sample, a cluster alignment method [27] is employed to analyze the atomic structure of the $Al_{90}Tb_{10}$ glass. We show that more than 80% the first-shell Tb-centered clusters have the '3661' or '15551' SRO motifs. The '3661' motif has a top triangle, two consecutive hexagons in the middle and a single atom at the bottom, while the '15551' motif contains three consecutive pentagonal rings around Tb on the first shell and capped by an atom on the top and bottom ring respectively. The cluster alignment analysis also shows almost all of Tb-centered '3661' and '15551' SRO clusters exhibit MRO extended out to third coordination shells and can be identified as the Bergman-type superclusters. Such a Bergman-type MRO is consistent with a highly percolated network formed by the '3661' and '15551' SRO clusters mainly by vertex- or face-sharing of the clusters throughout the glass sample. Using genetic algorithm together with the neural network interatomic potential, we also performed crystal structure searches at the compositions nearby $Al_{90}Tb_{10}$. Four low-energy metastable structures at the composition around $Al_{90}Tb_{10}$ have been revealed. It is interesting to note that these metastable crystalline structures exhibit the same '3661'or '15551' SRO structures around the Tb center atoms, respectively, as observed in the glass $Al_{90}Tb_{10}$. However, their MRO topologies are different. While the crystalline $Al_{38}Tb_4$ and $Al_{35}Tb_4$ structures with the '3661' SRO exhibit a MRO very similar to that observed in the glass phase, the other $Al_{38}Tb_4$ and $Al_{35}Tb_4$ structures with the '15551' SRO has the atomic parking in the second and third shell around the Tb centers very different from that of the Bergman-type MRO. These results from our machine learning potential MD simulations and genetic algorithm crystal structure predictions suggest that there is strong competition between glass formation and crystal nucleation and growth in this system, which would be responsible for its marginal GFA.

## II. Simulation Details

While MD simulation is a very powerful tool for investigating the structural and chemical orders in condensed systems at the atomic level, the reliability of the predictions from MD heavily relies on the accuracy and transferability of the interatomic potential used in the MD simulation. A*b initio* molecular dynamics (AIMD) can calculate the energies and interatomic forces at very high accuracy, but heavy computational cost of AIMD limits its applications to small number of atoms and short MD simulation time. On the other hand, MD simulations using classical interatomic potentials can handle much larger number of atoms and much longer simulation time, but the accuracy of empirical potentials used in the MD simulations is an issue of great concern.

Recently, we have developed an interatomic potential for Al-Tb system using a neural network deep machine learning method [28-31]. The data set used in the machine learning to train the interatomic potential is composed of snapshots of liquid $Al_{90}Tb_{10}$, pure Al and pure Tb liquids, and crystal structures of fcc Al, hcp Tb, $Al_{17}Tb_2$, $Al_4Tb$, $Al_3Tb$, $Al_2Tb$, $AlTb$, $Al_2Tb_3$, $Al_2Tb$ and $AlTb_3$. The trained neural network model reproduces well the *ab initio* results of energies and forces in the training data set. For more details of our neural network interatomic potential for Al-Tb alloys, we refer the readers to our previous publication [31]. Our previous study [31] showed that MD simulations using the obtained deep potential reproduce accurately the AIMD results of $Al_{90}Tb_{10}$ liquid. The success in the studies of $Al_{90}Tb_{10}$ liquid provides us with a promising opportunity to study the short- to medium-range orders in Al-Tb metallic glass at an accuracy comparable to AIMD while computationally much faster.

The LAMMPS package [32] is used to performed MD simulations to study the structures and properties of $Al_{90}Tb_{10}$ glass, utilizing the developed neural network potential. Our MD simulation cell contains 5000 atoms (4500 Al and 500 Tb) with periodic boundary conditions. An isothermal-isobaric ensemble and a Nose-Hoover thermostat are used in the simulations [33, 34]. The MD time step for the simulation is 2.5fs. First the liquid $Al_{90}Tb_{10}$ sample is equilibrated at 2000K for 1 ns and then is continuously quenched down to 300K at cooling rates of $10^{13}$, $10^{12}$ and $10^{11}$ K/s, respectively. Then atomic positions averaged over 500 ps at 300K are used for the structure analysis so that the effect of atomic thermal motions on the structure order analysis is minimized.

The potential energies $E-3k_BT$ as the function of temperature for the $Al_{90}Tb_{10}$ samples with different cooling rates obtained from our MD simulation are shown in Fig. S1 in Supplementary Materials. It can be seen that the $Al_{90}Tb_{10}$ liquid changes to glass around 900K inferred by the change in the slope of potential energy curve. This temperature is close to the eutectic temperature near the composition of 10 at. % Tb from the Al-Tb phase diagram as indicated by the arrow in Fig. S1. Although the energy curve does not have an abrupt change like crystalline transition, the rapid potential energy drop upon cooling implies that some strong order structures with much lower potential energy emerge in the sample during the cooling process. In this paper, we will use cluster alignment method [27] to investigate short- to medium-range structural order in the glass $Al_{90}Tb_{10}$ sample obtained from our MD simulations at the cooling rate of $10^{11}$ K/s.

### III. Results and discussions

#### A. Structure factors and pair correlations functions

Before analyzing the local structure order in the $Al_{90}Tb_{10}$ glass by cluster alignment analysis, we demonstrate the accuracy of the glass model obtained from our MD simulation by comparing the calculated structure factors S(q) with that from experimental XRD measurement [25]. As one can see from Fig. 1(a), the structure factor calculated from our MD simulation glass sample at 300 K agrees well with experimental data. Even the pre-peak around 1.3 Å$^{-1}$ observed in experiment is captured correctly in our simulation glass sample. It can be seen that with decreasing cooling rate the intensity of the pre-peak and the main peak in our calculated S(q) is enhanced. We note that in the experimental S(q), there is a small peak in the valley between the first and second main peaks. Although our simulation results do not display a clear peak there, a shoulder feature around this peak position can be seen.

In addition to the total structure factor, the partial structure factors can be easily obtained by MD simulations, which enables us to investigate the contributor of pre-peak in S(q). Fig. 1(b)-(d) show the weighted partial structure factors for Al-Al, Al-Tb, and Tb-Tb pairs (i.e., $\omega_{AlAl}S_{AlAl}(q), \omega_{AlTb}S_{AlTb}(q)$ and $\omega_{TbTb}S_{TbTb}(q)$) respectively. Here the definition of weighted factor $\omega_{\alpha\beta}$ and partial structure factor $S_{\alpha\beta}(q)$ follow those in ref. 23. It can be seen that around the pre-peak position, the weighted Al-Tb structure factor has a valley and the weighted Al-Al structure factor has only very small intensity. By contrast, the weighted Tb-Tb structure factor has a significant peak around pre-peak position which is enhanced as cooling rate decreased, as shown in Fig. 1(d). Because the total structure factor is sum of the weighted partial structure factors ($S(q) = \omega_{AlAl}S_{AlAl}(q) + \omega_{AlTb}S_{AlTb}(q) + \omega_{TbTb}S_{TbTb}(q)$), our results indicate that the pre-peak in S(q) of Al$_{90}$Tb$_{10}$ glass originates from Tb-Tb correlations. Fig. 1(e)-(g) show the partial pair correlation functions (PPCFs) for Al-Al, Al-Tb, and Tb-Tb, respectively. In the Tb-Tb PPCF, the heights of the peaks corresponding to second and third coordination shell is higher than that of the first shell and the difference increases as cooling rate reduced, which implies that the order extended to medium-range in Tb-centered cluster is crucial for understanding of glass $Al_{90}Tb_{10}$ structure. Our cluster alignment analysis presented below provides more detail insights showing that the atomic structure order in $Al_{90}Tb_{10}$ glass is dominated by Tb-Tb spatial correlations.

### B. Dominant SRO motifs

We first investigate the SRO in the $Al_{90}Tb_{10}$ glass sample from our MD simulation at the cooling rate of $10^{11}$ K/s, using the cluster-template alignment method [27]. Here we concentrate our analysis on the Tb-centered clusters since the structure feature of $Al_{90}Tb_{10}$ glass is most likely to be determined by Tb-centered clusters. The analysis for the Al-centered clusters is given in the Supplementary Materials (see Figs. S2 –S5). The templates used for the alignment analysis of the Tb-centered clusters and the distribution of the alignment score with respect to the corresponding templates are shown in Fig. 2. In addition to the '3661' and '15551' clusters discussed above, the templates also include the first shell Tb-centered clusters from the known crystalline $Al_{17}Tb_2$, $Al_4Tb$, $Al_{11}RE_3$, and $Al_3Tb$ structures. The crystal structure of $Al_{11}Sm_3$ has two different motifs around the RE atoms which are denoted by $Al_{11}RE_3$-1 and $Al_{11}RE_3$-2 respectively. The alignment score from our cluster alignment analysis reflects the structural similarity between the Tb-centered clusters in the glass sample to the aligned template structure. The smaller the score value, the more similar is the cluster extracted from the glass sample to the corresponding template. If we take a cutoff score of 0.16 to assign the clusters to different template motif, the population of different types of clusters in the glass sample can be quantified. We note that if the cluster has alignment score lower than cutoff for more than one templates, then the one with smallest score will be assigned to the cluster.

As shown in Fig. 2, the alignment scores of '3661' and '15551' motifs have large portion below the cutoff score of 0.16, while the fraction of other motifs below the cutoff score can be neglected in the $Al_{90}Tb_{10}$ glass. The '3661' and '15551' motifs have been shown to be the low-energy motifs and frequently show up in genetic algorithm (GA) search for crystal structures in Al-Sm system [35]. From Fig. 2(b), it can be seen that about 82% of Tb-centered clusters are identified as one of these two motifs in the glass sample at cooling rate of $10^{11}$ K/s, among them the fraction of '3661' cluster (~57%) is about twice of '15551' cluster (~25%). We note that the best template from the crystalline structures that fits to the clusters in the glass sample is that from the $Al_4Tb$ crystalline structures, which captures about 5% of the Tb-centered clusters with marginal alignment scores. Based on our cluster alignment results, it suggests that $Al_{90}Tb_{10}$ glass has pronounced ordering within the first coordination shell around the Tb atoms.

### C. MRO in $Al_{90}Tb_{10}$ glass

The development of local order beyond the first atomic shell to the second and third atomic coordination shells of the clusters is also an important key factor to understand the phase selection in rapid solidification process. In order to demonstrate such MRO up to the second and third atomic coordination shell, the clusters identified as Tb-centered '3661' and '15551' SROs from the analysis shown in Fig. 2 are extended to contain about 70 atoms. From PPCFs shown in Fig. 1(e)-(g), a Tb-centered cluster containing about 70 atoms will include the atoms up to third atomic shell (radius < 7Å). Then each set of these super clusters are overlapped respectively by shifting their center atoms to the same origin while keeping the orientations of the clusters same as those have the best SRO alignment with the template. The atomic distributions in these two types of superclusters after such an alignment procedure are shown in Fig. 3 and Fig. 4 respectively. The strong SRO in the first atomic shell ('3661' in Fig. 3 and '15551' in Fig. 4) can be well seen. The structural order up to the third cluster shell can also be seen although the order is not as strong as that in the first atomic shell. It is interesting to note that the first shell atoms in both the Tb-centered '3661' and '15551' superclusters are mainly Al atoms (red), except the bottom vertex site in the first shell of the '3661' cluster, as one can see from the left column in Fig. 3 and Fig. 4 respectively. On the other hand, the order of the Tb atoms in the second and third atomic shells are much stronger in both '3661' and '15551' superclusters as can be seen from the middle and right columns in Fig. 3 and Fig. 4 respectively.

In order to show the structural order of the superclusters more clearly, a Gaussian smearing scheme [36] is used to convert the spatial atomic distributions shown in the top two rows of Fig. 3 and Fig. 4 into atomic density distribution in the three-dimensional space. The total atomic density (by adding together the contributions from both Al and Tb atoms) for the first, second, and third shell of the '3661' and '15551' superclusters are shown in bottom row of Fig. 3 and Fig. 4 respectively. The structure at each shell of the supercluster is represented by a cage whose vertexes are located at the centers of the atomic density blobs in each shell (first shell in gray, second shell in red and third shell in green). The plot reveals both the '3661' and '15551' superclusters have a Bergman-type structure, i.e., the atoms in the second and third shells tend to cap the faces of the inner cages, which is similar to the case of Bergman supercluster in $Cu_{64.5}Zr_{35.5}$ metallic glass [36]. Here, the atoms in Bergman type positions can be viewed as a close atomic packing outward from the surface of the first shell around the Tb centers. We noted that the order of the second shell Bergman packing is stronger at the bottom hex cap of the '3661' cluster and at the top and bottom pentagonal caps of the '15551' cluster. Beside the second shell, the third shell atoms also prefer to occupy the face-caped positions with respect to the second shell to form the Bergman-type MROs (denoted by the green bonds) in both '3661' and '15551' superclusters.

To further quantify the degree of Bergman-type MRO in the glass sample, the cluster-template alignment analysis is also performed using the Bergman-type 42-atoms '3661' and 48-atoms '15551' superclusters up to the second shell as templates (see the template structures and alignment scores in the Fig. S6 of Supplementary Materials). The size of the '3661' and '15551' clusters identified from SRO analysis are extended out to the second shell and then these superclusters are extracted from the glass sample to align against the Bergman-type '3661' and '15551' templates respectively. As one can see from Fig. S6, the alignment scores for almost all the '3661' and '15551' superclusters in the sample are smaller than 0.2 which is an acceptable score value to assign the aligned supercluster to the template. Therefore, the alignment results show that most of Tb-centered clusters not only have pronounced orders in their first shell but also such SROs extend to the second and third shells forming Bergman-type superclusters.

Besides the cluster alignment analysis, an alternative way to investigate MRO in metallic glass is to examine the spatial correlation of the SRO clusters to see how a network is formed by these clusters in the glass. In our analysis, two first-shell '3661' and '15551' SRO clusters are considered as connected if they share at least one atom. There are four different types of connections between two adjacent clusters, i.e., vertex sharing, edge sharing, face sharing and inter-penetrating. Fig. 5(a) show the spatial network formed by the connections of the '3661' and '15551' clusters in our sample at T=300K prepared with the cooling rate of $10^{11}$ K/s from the liquid. It can be seen that at all the Tb centered '3661' and '15551' clusters are connected forming a well percolated network, which can be clearly demonstrated by calculating the maximal size of network as shown in the Supplementary Materials Fig. S7. We found that all '3661' clusters in the sample are connected by themselves, while about 80% '15551' clusters are connected among themselves. However, the spatial distribution of the '3661' and '15551' clusters are highly intermixed with each other. The analysis shows that that all the '15551' clusters have connections to at least one '3661' cluster. The cluster network is composed of about 43% of vertex and 39% of facing sharing connections, while cluster interpenetrating connection is only about 6%, as shown in the Supplementary Materials Fig. S8.

In order to see the connection between the descriptions of MRO in terms of SRO clusters network and in terms of Bergman-type superclusters packing, we analyzed the distances between the center Tb atoms of the adjacent '3661' or/and '15551' clusters according to the different type of cluster sharing topologies and plot them in Fig. 5(b). More detailed information about the distances between two adjacent SRO centers can be also found in Fig. S9 of Supplementary Materials. By comparing these distance distributions with the Tb-Tb PPCF shown in Fig. 5 (c), it can be seen that the Tb centers belong to face sharing clusters are in the second shell (indicated by red arrow) while those belong to vertex sharing clusters are in the third shell of the Bergman-type superclusters (indicated by blue arrow). To see this point more clearly, in Fig. 5(d) we plotted the atoms in a typical part of the '3661' network. For the face sharing cluster connections, it can be seen that the two Tb centers and three Al atoms of the shared triangle form a nearly ideal triangular bipyramid. As shown in Fig. 5(e), one of vertex in the triangular bipyramid is just located at the Bergman-type position in the second shell of '3661' cluster. The distance between these two Tb centers is $D_2 \sim 5.5$Å which is consistent with the results of distance analysis between two face sharing '3661' clusters. Thus, the pronounced ordering of the second shell implies that the network with face sharing connections will develop along the directions of Bergman-type sites. Similar to the case of face sharing connections, the network with vertex sharing connection will grow along the sites of big '3661' supercluster in the third shell.

The structural order of the face and vertex connections in the network formed by the Tb-center clusters is also correlated with the chemical order of Tb atom in the first, second, and third coordination shell for the '3661' and '15551' superclusters as plotted in Fig. 3 and 4, respectively. Since almost all the Tb atoms (∼82%) are the centers of '3661' or '15551' clusters, most Tb atoms in superclusters represents a Tb-centered '3661' or '15551' cluster. It can be seen that in the first shell only a small amount of Tb atoms are present. However, similar to the case in $Al_{90}Sm_{10}$ metallic glass, the site '1' of '3661' motif has much larger possibility of Tb occupancy than the other sites, implying the interpenetrating two '3661' SRO is mainly formed through this site. For the second and third shell, the possibility of Tb occupancy increases significantly, which is consistent with the distribution of distance for two SRO centers (see Fig. 5(b)) or PPCF of Tb-Tb (see Fig.1 (g)). The possibility of Tb atoms in all the second and third shell sites of the Bergman cluster is similar, indicating that the face and vertex sharing connections develop nearly equally along all the Bergman-type directions. More details of the statistics of the SRO cluster connection in the first, second, and third shell of the Bergman superclusters shown in the Fig. S10 of the Supplementary Materials also support our conclusion about the relation between the SRO cluster network and the Bergman superclusters in $Al_{90}Tb_{10}$ glass. From Fig. 3 and Fig. 4 we can also see that the spatial distribution of the Tb atoms is also highly angular correlated. i.e., the Tb-center atoms in the '3661' and '15551' clusters will extend along the directions of Bergman-type sites.

**D. Connection to the order in competing crystalline phases**

In order to understand how the SRO and MRO observed in the glass phase of $Al_{90}Tb_{10}$ are related to the structures of possible competing crystalline phases in the same composition region, we performed crystal structure search around the composition of $Al_{90}Tb_{10}$ by genetic algorithm (GA) using the same neural network interatomic potential employed in our MD simulation of glass. Two low-energy crystalline structures at each composition of $Al_{38}Tb_4$ and $Al_{35}Tb_4$ respectively have been revealed from the GA search and shown in Fig. 6. The energetic stability of these four crystalline structures has also been verified by first-principles calculations. One of the $Al_{38}Tb_4$ structures has the '3661' cluster as its building block while another structure has the '15551' motif as shown in Fig. 6 (a) and (d), respectively. Similarly, the two $Al_{35}Tb_4$ crystalline structures also have the '3661' and '15551' SRO motifs respectively as shown in Fig. 6 (b) and (e). All the four crystalline structures have the favorable formation energies. From the first-principles DFT calculations, the formation energies of the two $Al_{38}Tb_4$ structures are 5.04 eV/f.u. while that pt the two $Al_{35}Tb_4$ structures are 4.68 eV/f.u. and 5.37 eV/f.u., respectively. These formation energies are only about 40 to 60 meV/atom above the corresponding convex hull of the Al-Tb binaries. Therefore, these four structures, with the same SRO motifs as observed in the glass, are energetically very competitive in phase selection against glass formation around the $Al_{90}Tb_{10}$ composition. It is interesting to note that while these four crystalline structures exhibit the SRO as those in the $Al_{90}Tb_{10}$ glass, their MRO packing around the Tb centers would be different. While the atomic packing on the second and third shells in the crystalline structures with the '3661' motif is very similar to the Bergman-type supercluster observed in the glass (there are only some subtle difference at the top and bottom region of the supercluster as shown in Fig. 6 (b) and (c)), the MRO packing beyond the first shell in the two crystalline structures with the '15551' motif is very different from the Bergman type of packing, as one can see from Fig. 6(d)-(f).

IV. **Summary**

In summary, the short-range and medium-range structural orders in $Al_{90}Tb_{10}$ metallic glass have been studied based on the atomistic structures from MD simulations using an interatomic potential developed by neural network deep learning method. The calculated structure factor agrees well with the XRD experimental data, especially for the pre-peak position and intensity. With decreasing cooling rate, the pre-peak is enhanced and the partial structure factors show that the Tb-Tb correlation is responsible for the pre-peak in S(q).

Using the cluster alignment method, we identify the '3661' and '15551' motifs are the predominant SRO clusters in the $Al_{90}Tb_{10}$ glass. The MRO structures up to the second and third coordination shells of Tb-centered cluster are also studied. It is found that the Bergman-type packing in the second shell of '3661' and '15551' clusters. The order is much more pronounced at the Bergman-type positions corresponding to the geometric perfect hex and pentagonal cap in '3661' and '15551' motifs, implying that the geometry of first shell order is important for the MRO development. In the third shell, there is still observable ordering although it is less pronounced than that in the second shell. Cluster alignment using the MRO Bergman-type '3661' and '15551' supercluster as templates also show that almost all of Tb-centered clusters are Bergman-type '3661' or '15551' superclusters.

We also study the MRO from the cluster network point of view. We show that the predominate '3661' and '15551' clusters are highly inter-mixed and form a well percolated network throughout the glassy sample. The connections between the clusters in the network are mainly face and vertex sharing types, while there is only a small portion of connections are interpenetrating type. This type of cluster network topology is consistent with the Bergman-type of MRO from the supercluster atomic packing point of view. We believe the degree of such Bergman MRO would be further developed at lower cooling rate during the glass sample preparation.

Moreover, we also show that low-energy metastable crystalline structures in the same composition range also exhibit the same '3661' and '15551' SRO, and even the same '3661' Bergman-type MRO. Such energetic competitive crystalline genes found in the glass phase would be the origin of marginal glass-forming ability of $Al_{90}Tb_{10}$.

**Acknowledgements**

Work at Ames Laboratory was supported by the U.S. Department of Energy (DOE), Office of Science, Basic Energy Sciences, Materials Science and Engineering Division including a grant of computer time at the National Energy Research Supercomputing Center (NERSC) in Berkeley. Ames Laboratory is operated for the U.S. DOE by Iowa State University under contract # DE-AC02-07CH11358. L. Tang and Z. J. Yang acknowledge the support by the National Natural Science Foundation of China (Grant Nos. 11304279 and 11104247). Z. J. Yang also acknowledges the Natural Science

Foundation of Zhejiang Province, China (Grant No. LY18E010007).**References**
[1] Y. He, S. J. Poon, G. J. Shiflet, Science 241, 1640 (1988).
[2] J. H. Perepezko and R. J. Hebert, JOM 54, 34 (2002).
[3] J. H. Perepezko, Progress in Materials Science 49, 263 (2004).
[4] K. F. Kelton, A. L. Greer, and K. Rajan, Journal of Non-Crystalline Solids, 317, Page vii (2003).
[5] X.-L. Wang, J. Almer, C. T. Liu, Y. D. Wang, J. K. Zhao, A. D. Stoica, D. R. Haeffner, and W. H. Wang, Phys. Rev. Lett. 91, 265501 (2003).
[6] C.Yildirim, M.Kutsal, R.T.Ott, M.F.Besser, M.J.Kramer, Y.E.Kalay, Materials & Design 112, 479 (2016).
[7] Eric T. Stromme, Hunter B. Henderson, Zachary C. Sims, Michael S. Kesler, David Weiss, Ryan T. Ott, Fanqiang Meng, Sam Kassoumeh, James Evangelista, Gerald Begley, and Orlando Rios, JOM 70, 866 (2018).
[8] A. Inoue, Prog. Mater. Sci. 43, 365 (1998).
[9] D.V Louzguine and A. Inoue, Journal of Non-Crystalline Solids 311, 281 (2002).
[10] D.V Louzguine and A. Inoue, Journal of Non-Crystalline Solids 352, 3903 (2006).
[11] J.H. Perepezko and R.J. Hebert, Journal of Metals 54, 34 (2002).
[12] J.H. Perepezko, R.J. Hebert and W.S. Tong, Intermetallics 10, 1079 (2002).
[13] W.G. Stratton, J. Hamann, J.H. Perepezko, P.M. Voyles, X. Mao and S.V. Khare, Appl. Phys. Lett., 86, 141910 (2005).
[14] R. Hebert, J.H. Perepezko, H. Rösner and G. Wilde, Scripta Mat., 54, 25 (2006).
[15] A. K. Gangopadhyay, T. K. Croat, K. F. Kelton, Acta Materialia 48, 4035 (2000).
[16] K.K.Sahua, N.A.Mauro, L.Longstreth-Spoor, D.Saha, Z.Nussinov, M.K.Miller, K.F.Kelton, Acta Materialia 58, 4199 (2010).
[17] A. Inoue, K. Ohtera, and T. Masumoto, Jpn. J. Appl. Phys. 27, L736 (1988).
[18] X. M. Li, Y. Wang, J. J. Yi, L. T. Kong, and J. F. Li, J. Alloy. Compd. 790, 626 (2019).
[19] H. B. Lou, Z. D. Zeng, F. Zhang, S. Y. Chen, P. Luo, X. H. Chen, Y. Ren, V. B. Prakapenka, C. Prescher, X. B. Zuo, T. Li, J. G. Wen, W. H. Wang, H. W. Sheng, and Q. S. Zeng, Nat. Commun. 11 (1), 9 (2020).
[20] L.Zhao, G.B.Bokas, J.H.Perepezko, and I.Szlufarska, Acta Materialia 142, 1 (2018).
[21] G.B.Bokas, L.Zhao, J.H.Perepezko, and I.Szlufarska, Scripta Materialia 124, 99 (2016).
[22] M. I. Mendelev, F. Zhang, Z. Ye, Y. Sun, M. C. Nguyen, S. R. Wilson, C. Z. Wang, and K. M. Ho, Modelling Simul. Mater. Sci. Eng. 23, 045013 (2015).
[23] Feng Zhang, Yang Sun, Zhuo Ye, Yue Zhang, Cai-Zhuang Wang, Mikhail I Mendelev, Ryan T Ott, Matthew J Kramer, Ze-Jun Ding, and Kai-Ming Ho, J. Phys.: Condens. Matter 27, 205701(2015).
[24] Yang Sun, Yue Zhang, Feng Zhang, Zhuo Ye, Zejun Ding, Cai-Zhuang Wang, and18

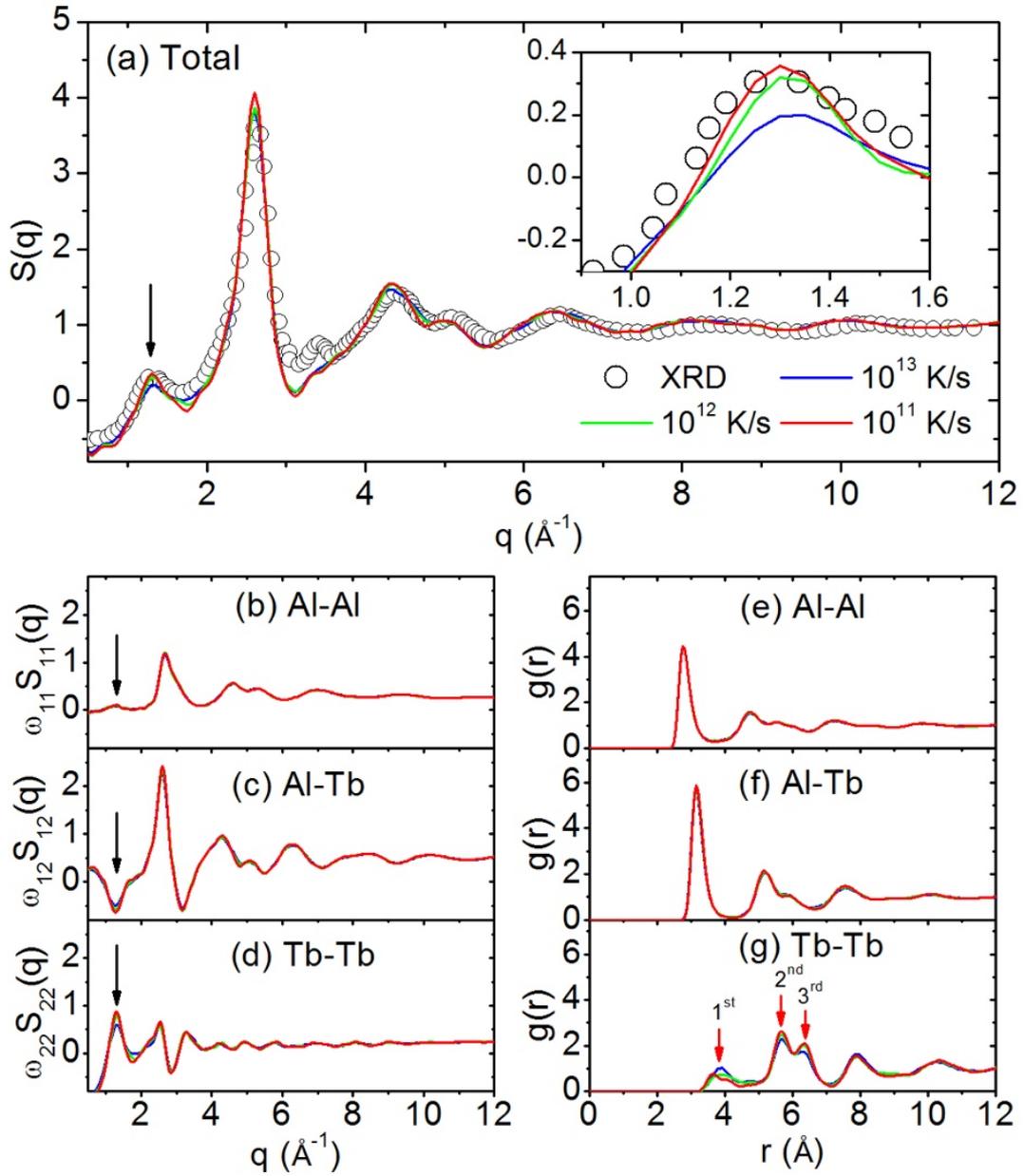

Fig. 1. (a) The comparisons of total structure factors from the MD simulations with the XRD data of amorphous $Al_{90}Tb_{10}$ at T=300K. The open circle is XRD data and the solid lines are S(q) obtained by the MD. The inset zoom-in the S(q) around the pre-peak structure. Figure (b)-(d) are the weighted partial structure factors of Al-Al, Al-Tb, and Tb-Tb, respectively. Here the black arrows indicate the position of pre-peak $q = 1.3$ Å. The weighted Tb-Tb structure factor has a significant peak at $q = 1.3$ Å, suggesting that the Tb-Tb correlation is responsible for the pre-peak of S(q). Figure (e)-(g) are the partial pair correlation functions of Al-Al, Al-Tb, and Tb-Tb, respectively. The red arrows denote the first, second, and third coordination shell of Tb-Tb PPCF.

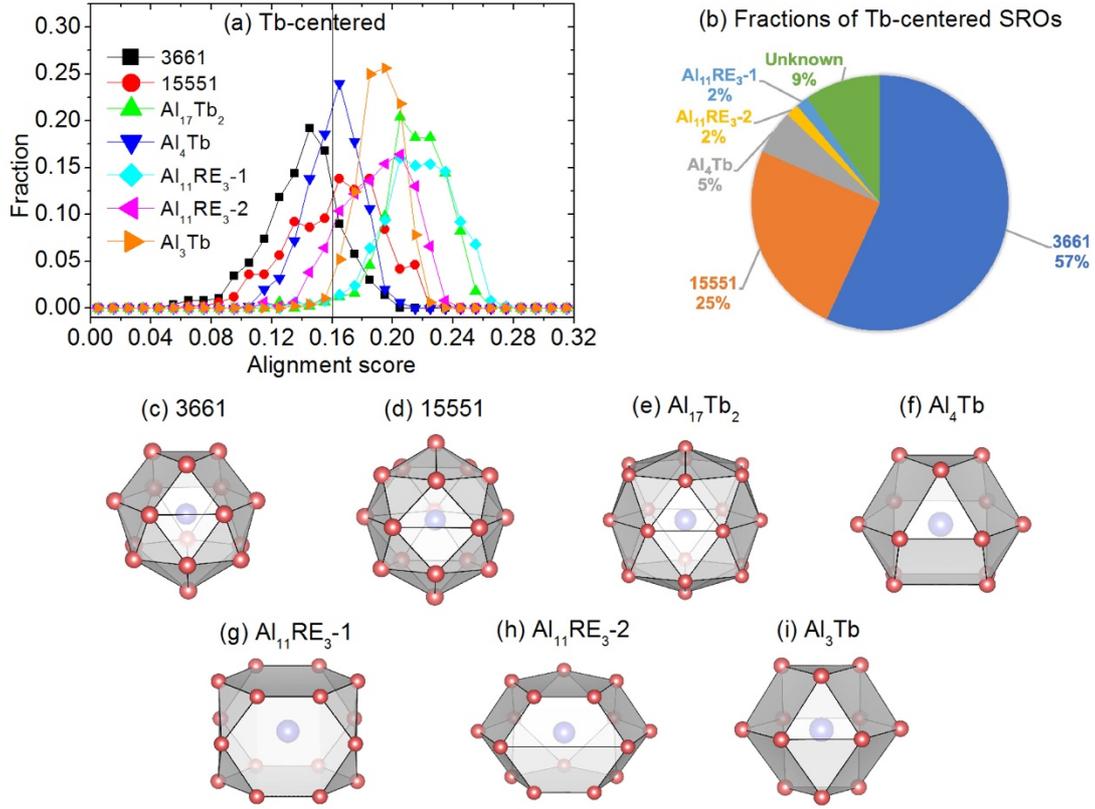

Fig. 2. (a) The distribution of alignment scores against various motifs for Tb-centered clusters in $Al_{90}Tb_{10}$ at T=300K quenched with cooling rate of $10^{11}$ K/s. The atomic structures of '3661' and '15551' template as well as those extracted from the known metastable crystal phases ($Al_{17}Tb_2$ and $Al_4Tb$), stable phase ($Al_3Tb$), and the crystal with $Al_{11}Sm_3$ structure (two motifs denoted by $Al_{11}RE_3$-1 and -2) are also shown in (c)-(i). (b) The fractions of Tb-centered SROs in the sample. The '3661' and '15551' motifs are dominant SROs (total fraction is about 82%). The total fraction of motifs from known crystal phases ($Al_{17}Tb_2$, $Al_4Tb$, $Al_3Tb$, $Al_{11}RE_3$) is about 9%. The fraction of rest clusters with unknown atomic structures is also about 9%.

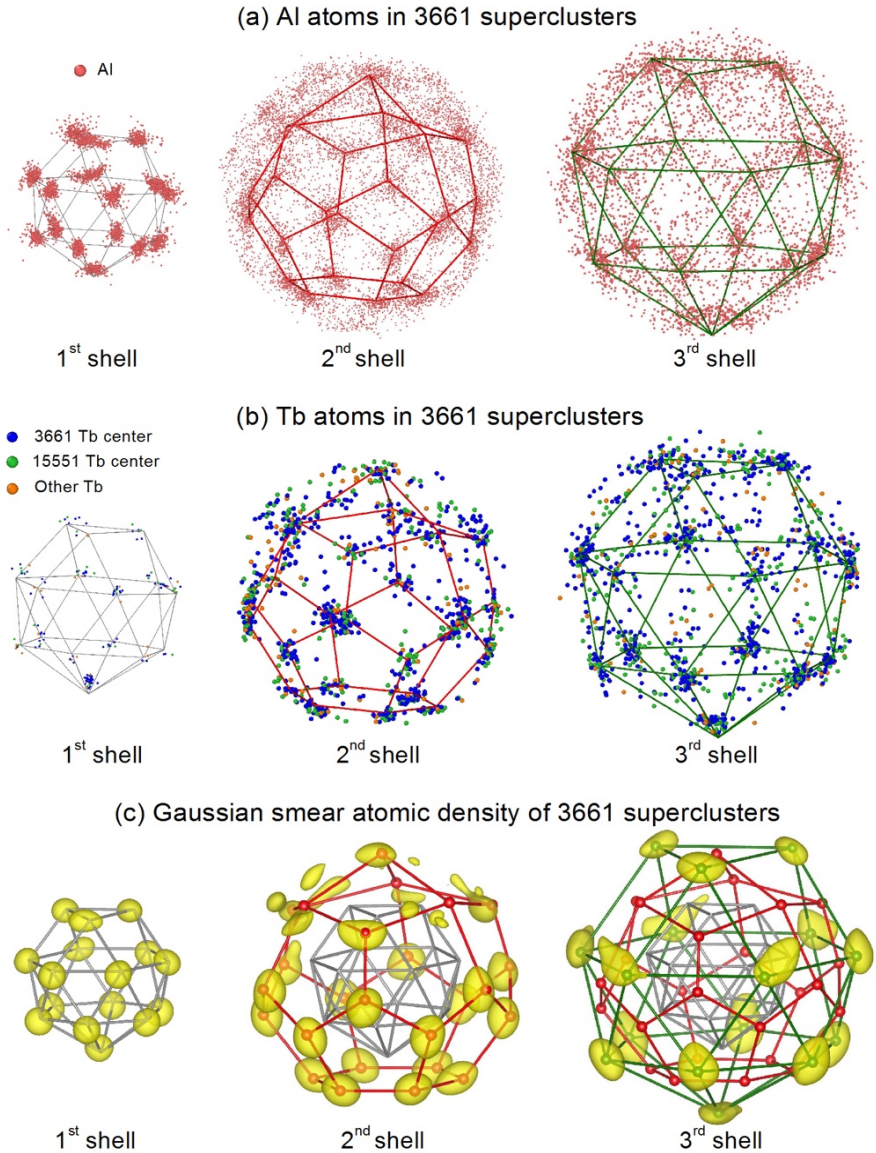

Fig. 3. The atomic distribution of (a) Al and (b) Tb in the first, second, and third coordination shell for '3661' superclusters. The red balls denote the Al atoms. The blue, green, and yellow balls respectively denote the Tb centers of '3661', '15551' SROs, and the other Tb atoms which are not the centers of dominant SROs. (c) The iso-surface of atomic density obtained by Gaussian smear method for '3661' superclusters. Here, the gray bonds are the '3661' SRO in the first shell. The red bonds are to demonstrate the Bergman-type supercluster in the second shell. The green bonds denote the supercluster of the third shell, which is a big '3661' cluster.

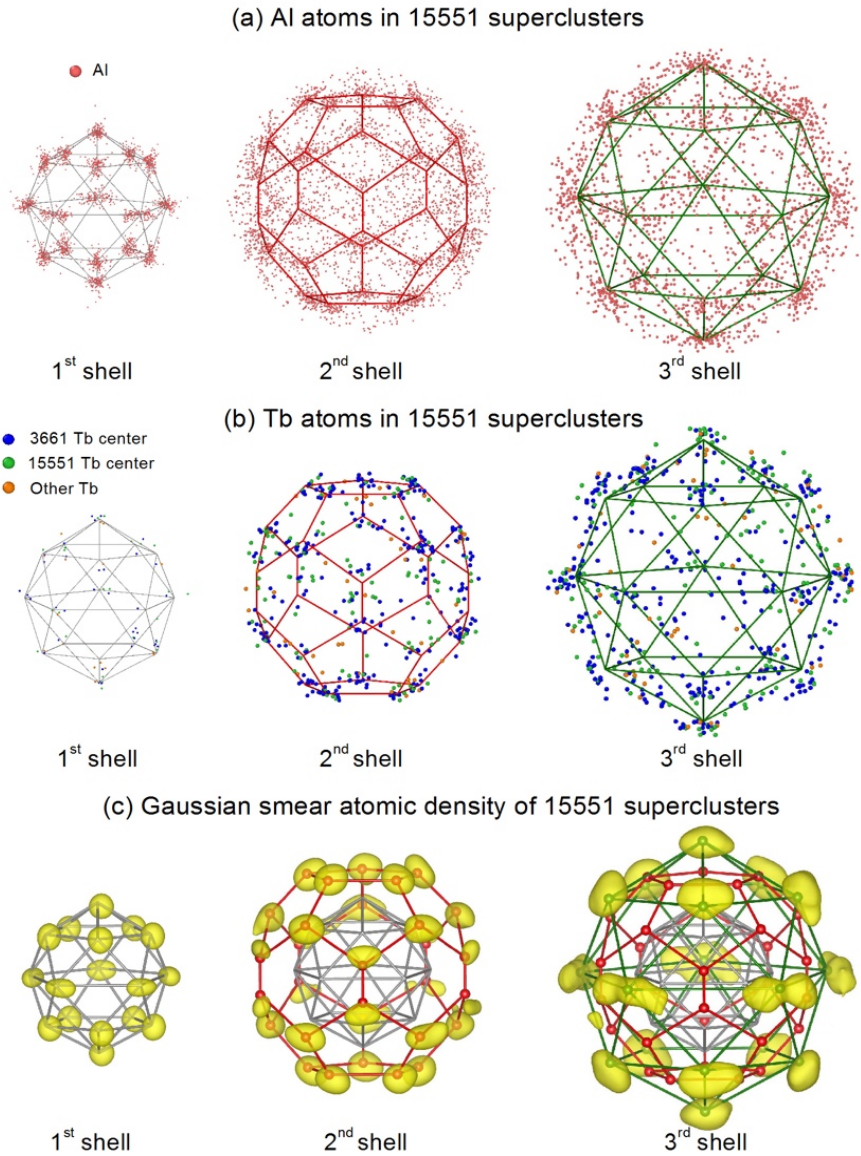

Fig. 4. The atomic distribution of (a) Al and (b) Tb in the first, second, and third coordination shell for '15551' superclusters. The red balls denote the Al atoms. The blue, green, and yellow balls respectively denote the Tb centers of '3661', '15551' SROs, and the other Tb atoms which are not the centers of dominant SROs. (c) The iso-surface of atomic density obtained by Gaussian smear method for '15551' superclusters. Here, the gray bonds are the '15551' SRO in the first shell. The red bonds are to demonstrate the Bergman-type supercluster in the second shell. The green bonds denote the supercluster of the third shell, which is a big '15551' cluster.

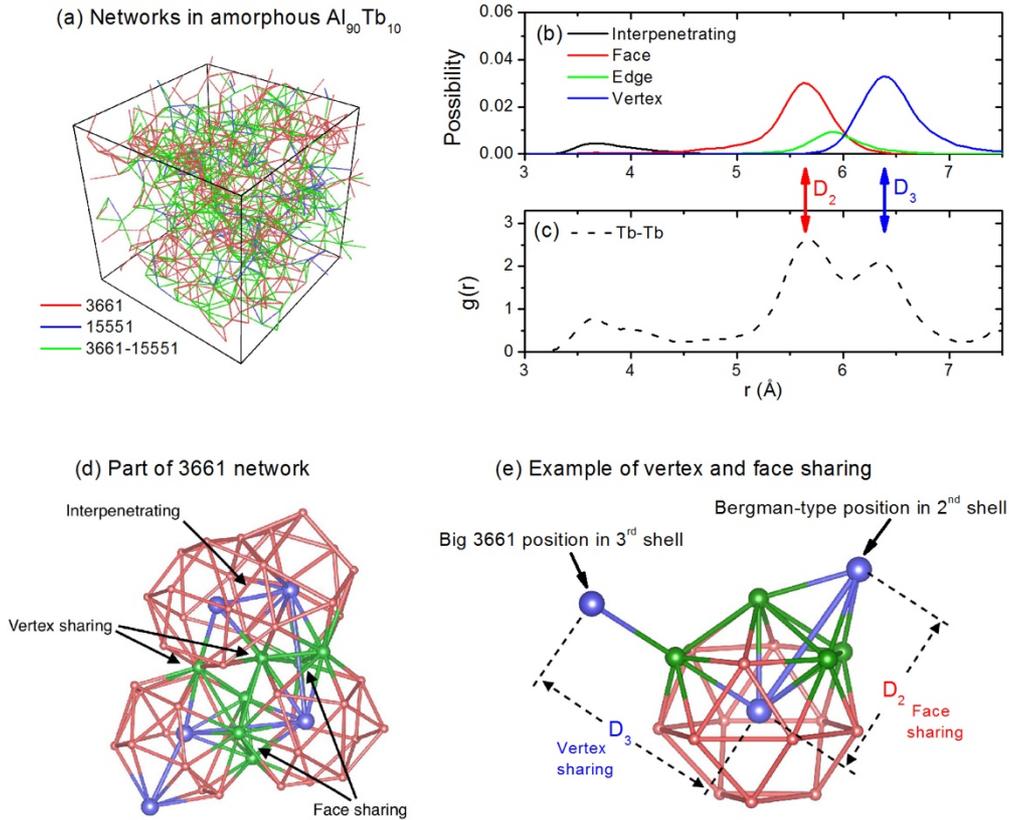

Fig. 5. (a) The network of '3661', '15551' clusters, and their interconnection ('3661'-'15551') in the sample. Here if any two SRO clusters share at least one atom, they are considered as the connected network and plotted with bonds. The two centers of clusters are connected in four types of connection mode, i.e., vertex, edge, face sharing and interpenetrating. (b) The distribution of distance between two adjacent SRO Tb centers with different connection types. (c) The PPCF of Tb-Tb. (d) A typical part of '3661' network with four '3661' clusters. The blue, red, and green balls are Tb, Al, and the shared Al atoms by two '3661' clusters, respectively. (e) The example of connections with vertex and face sharing. For clarity, only the center Tb atoms of adjacent '3661' clusters are shown. As shown in the right part of (e), the center Tb of face sharing just occupies the Bergman-type position of the second shell (The distance $D_2$ is demonstrated by the red arrow in (b) and (c)). The left part of (e) shows that the two Tb center atoms of vertex sharing are collinear with the shared Al atom. The Tb centers of vertex sharing just occupies the site of big '3661' supercluster (The distance $D_3$ is demonstrated by the blue arrow in (b) and (c)). In other words, it demonstrates that the Tb in the Bergman supercluster has face sharing with center Tb, meanwhile the Tb in the big '3661'(or '15551') supercluster has vertex sharing with center Tb.

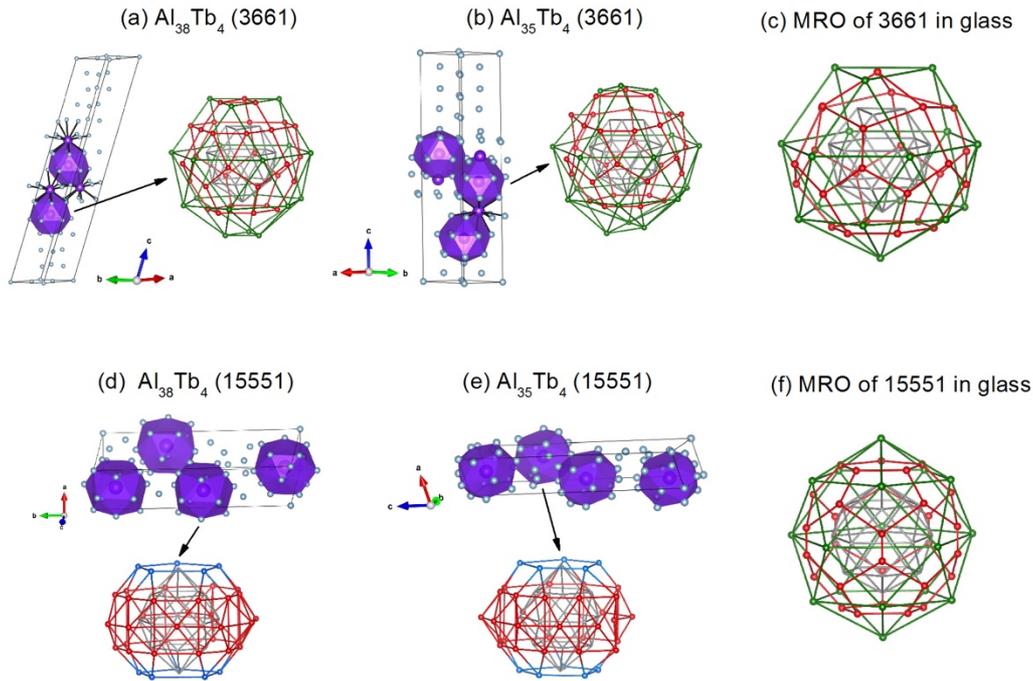

Fig. 6. Low-energy metastable crystalline structures of $Al_{38}Tb_4$ and $Al_{35}Tb_4$ from GA search using the neural network machine learning potential and confirmed by *ab initio* calculations. The SRO building block in the crystal (a) and (b) is the Tb-centered '3661' cluster while that in the crystal (d) and (e) is the Tb-centered '15551' cluster. These two types of clusters are the dominant SRO motifs observed in $Al_{90}Tb_{10}$ glass. While the atomic packing on the second and third shells around the '3661' first shell in the crystal (a) and (b) is very similar to the Bergman-type supercluster observed in the glass (except some subtle differences at the top and bottom region of the supercluster) as shown in (c), the atomic packing beyond the first shell of the '15551' MRO in the crystal structure (d) and (e) is very different from the Bergman type of packing, as one can see from (f). Here, the blue balls in (d) and (e) denote the only sites near the Bergman-type positions in crystal (d) and (e).